\documentclass[12pt,preprint]{aastex}

\newcommand{\galex}{{\it GALEX~}}

\shorttitle{UV Rising Flux}
\shortauthors{RICH ET AL.}

\received{2004 June 25}
\begin{document}

\title{Systematics of the Ultraviolet Rising Flux in a {\sl GALEX}/SDSS Sample
of Early-type Galaxies}

\author{
R.\ Michael Rich\altaffilmark{1},
Samir Salim\altaffilmark{1},
Jarle Brinchmann\altaffilmark{2,13},
St\'ephane Charlot\altaffilmark{2,12},
Mark Seibert\altaffilmark{4},
Guinevere Kauffmann\altaffilmark{2},
Young-Wook Lee\altaffilmark{6},
S.K.Yi\altaffilmark{6,14},
Tom A.\ Barlow\altaffilmark{4},
Luciana Bianchi\altaffilmark{5},Yong-Ik Byun\altaffilmark{6}, 
Jose Donas\altaffilmark{7},
Karl Forster\altaffilmark{4},
Peter G.\ Friedman\altaffilmark{4},
Timothy M.\ Heckman\altaffilmark{3},
Patrick N.\ Jelinsky\altaffilmark{8},
Barry F.\ Madore\altaffilmark{9,10},
Roger F.\ Malina\altaffilmark{7},
D.\ Christopher Martin\altaffilmark{4},
Bruno Milliard\altaffilmark{7},
Patrick Morrissey\altaffilmark{4},
Susan G.\ Neff\altaffilmark{11},
David Schiminovich\altaffilmark{4},
Oswald H.\ W.\ Siegmund\altaffilmark{8},
Todd Small\altaffilmark{4},
Alex S.\ Szalay\altaffilmark{3},
Barry Y.\ Welsh\altaffilmark{8}, and
Ted K.\ Wyder\altaffilmark{4}
}
\email{rmr@astro.ucla.edu}

\altaffiltext{1}{Dept. of Physics and Astronomy, UCLA, Los
Angeles, CA 90095-1547}

\altaffiltext{2}{Max-Planck Institut f\"ur Astrophysik, 
D-85748 Garching, Germany}

\altaffiltext{3}{Department of Physics and Astronomy, The Johns Hopkins
University, Homewood Campus, Baltimore, MD 21218}

\altaffiltext{4}{California Institute of Technology, MC 405-47, 1200 East
California Boulevard, Pasadena, CA 91125}

\altaffiltext{5}{Center for Astrophysical Sciences, The Johns Hopkins
University, 3400 N. Charles St., Baltimore, MD 21218}

\altaffiltext{6}{Center for Space Astrophysics, Yonsei University, Seoul
120-749, Korea}

\altaffiltext{7}{Laboratoire d'Astrophysique de Marseille, BP 8, Traverse
du Siphon, 13376 Marseille Cedex 12, France}

\altaffiltext{8}{Space Sciences Laboratory, University of California at
Berkeley, 601 Campbell Hall, Berkeley, CA 94720}

\altaffiltext{9}{Observatories of the Carnegie Institution of Washington,
813 Santa Barbara St., Pasadena, CA 91101}

\altaffiltext{10}{NASA/IPAC Extragalactic Database, California Institute
of Technology, Pasadena, CA}

\altaffiltext{11}{Laboratory for Astronomy and Solar Physics, NASA Goddard
Space Flight Center, Greenbelt, MD 20771}

\altaffiltext{12}{Institut d'Astrophysique de Paris, CNRS,
98 bis boulevard Arago, F-75014 Paris, France}

\altaffiltext{13}{Centro de Astrofisica da Universidade do Porto, Rua 
das Estrelas, 4150-762 Porto, Portuga}

\altaffiltext{14}{Department of Physics, University of Oxford, Oxford 
OX1 3RH, UK}

\begin{abstract}
We present ultraviolet photometry for a sample of morphologically early-type
galaxies selected by matching the Sloan Digital Sky Survey
Data Release 1 with the \galex Medium and All-sky
Imaging Surveys.   We obtain a working sample of 1032 early-type 
galaxies with \galex
FUV detections, SDSS spectroscopy,
and $z<0.2$.  Using the SDSS spectra to identify galaxies with even
weak star formation or evidence of AGN, and further removing galaxies with
any evidence of non early-type morphology, we derive a final sample of 172 red
quiescent early-type galaxies.   We find that the $FUV-r$ color has a full range
of 5 mag.  Plotting against the $FUV-r$ color
the metallicity sensitive Lick $\rm Mg_2$ and D4000 indices, and the stellar velocity dispersion,
we find no correlation between our measurement of UV rising flux, and any parameter
sensitive to metallicity.  

\end{abstract}


\section{Introduction}

The unexpected measurement of ultraviolet light in the metal rich populations
of bulges and spheroids was one of the early important discoveries in space astronomy \citep{code69}.  Initial suggestions that the ultraviolet rising
flux (or UVX) was caused
by low level star formation or nuclear activity
gave way to the current view, that evolved stars are
responsible for the UVX.  
While not arising from recent
star formation,  the stellar population responsible for the UVX in early
type galaxies has eluded convincing identification among the resolved
stars in globular clusters and other nearby populations.  Early theoretical
work aimed at solving the mystery (e.g., \citealt{greg90}) 
favors evolved stars as the source,
with helium burning stars 
being both luminous enough, and sufficiently long lived, to be plausible
candidates.    Recent observations of nearby galaxies support this view.
Modeling integrated spectra of nearby UVX galaxies,
\citet{brown97} argue against post-AGB stars as the dominant contributor.
Direct imaging of M32 \citep{brown00} finds fewer than
expected luminous UV stars (and a striking lack of post-AGB stars) firming
support for the extreme horizontal branch (EHB) hypothesis.
Recent reviews considering the problem of the UVX
are given in \citet{oconnell99}, \citet{greg-ren99}, and \citet{brown03}.

One of the core observational results is the correlation of the UV rising flux with metallicity and
velocity dispersion (\citealt{b88};BBBFL).  BBBFL show that the nuclear
populations of 24 early-type galaxies with optical spectra indicating
no signs of AGN or star formation
exhibit a strong correlation
between UVX (in the form of a $1550-V$ color) and the Lick
$\rm Mg_2$ metallicity index, a result which strongly influenced
theoretical efforts to explain the UVX.
The discovery of very blue horizontal branch
stars in the metal rich globular clusters NGC 6388 and 6441 \citep{rich97} 
shows that UV bright stars are present even in some old
metal rich populations.

The presence of a correlation between UVX and independent measurable
parameters not only gives a clue to the source of the UVX, but also
could be exploited to gain insight into the evolution of the UVX with
lookback time.  If the source of the UVX is hot horizontal branch
stars, theoretical models (e.g. \citealt{yi99}) expect a 
rapid decline as a function of lookback time, because only within 
the last few Gyr have the masses of old turnoff stars been low enough
that the helium burning progeny of RGB tip stars would have envelopes
thin enough to make the stars UV bright.  To make possible the study of the UVX
in large samples, or distant galaxies, we seek correlations in properties
measurable from integrated light.

Recent observations have begun to weaken the assertion of
a strong primary correlation between UVX and metallicity.
\citet{ohl98} use the film-based imaging of the {\it Ultraviolet
Imaging Telescope} \citep{ocon92} to find far-UV color gradients in early-type
galaxies, in the sense of redder colors in the optically bluer, outer parts
of ellipticals.  However, this study finds a range in color gradients
and no clear trend with the Lick $\rm Mg_2$ index.  \citet{dehar02} form
a color using $m(2000)-V$ as measured from the balloon-borne {\it FOCA}
mission \citep{donas87} and plot this against the $\rm Mg_2$ index
for early-type galaxies.  They find no correlation between $m(2000)-V$ and
$\rm Mg_2$ for the early-type galaxies in their sample.  However, they
do not spectroscopically screen against galaxies with AGN and weak
star formation and, most significantly, the UVX dominates
{\it blueward} of 2000\AA, in the satellite ultraviolet.

The successful launch of the {\sl Galaxy Evolution Explorer} has made possible
the construction of a significantly larger sample of early-type galaxies
with which to probe the behavior of UVX 
than was possible for BBBFL. The observations above the atmosphere 
permit measurement of the UVX in the optimal window where it is most
prominent,
from $1200-1800$\AA.  The satellite and its operations are described in this
volume \citep{martin04,mor04}.

In this {\sl Letter} we explore the relationship
between the UVX (in the form of integrated light $FUV-r$ 
and metallicity) for a sample of galaxies considerably
more distant than was considered in BBBFL, but nonetheless 
early-type galaxies in the relatively local Universe $(z<0.2)$.
We use the $FUV-r$ color as the best proxy for the $1550-V$
color of BBBFL, given that SDSS $r$ is generally measured with
higher S/N and has a smaller K-correction than SDSS $g$.

\section{Data and Sample}
We aim to  combine GALEX and SDSS data
for a sample of early-type galaxies with no optical signs of star formation,
nuclear activity, or disturbed morphology.  We require that the
galaxies have SDSS spectra with high enough S/N to assure
neither detectable $\rm H\alpha$ nor K+A type spectra will
enter this sample.    These ``quiescent'' early-type 
(QE) galaxies are selected based on morphology and spectroscopy.
We begin by selecting candidates 
from the SDSS DR1 spectroscopic sample that are
consistent with early-type morphology and basic spectral shape. We
draw our sample from objects photometrically classified as galaxies,
and use essentially the same criteria used by \cite{bern}: a)
$(R_{90}/R_{50})_{i} > 2.5$, the high ratio of radii capturing 90 and
50\% of the Petrosian flux (also known as concentration) ensures that
most early-type galaxies are selected (but with sizable late-type
contamination); b) $\ln {\rm lDev}_r > \ln {\rm lExp}_r + 0.03$
selects objects whose likelihood of a deVaucouleurs ($R^{1/4}$)
profile in $r$-band is 3\% higher than the exponential profile; c)
${\rm eClass} > -0.1$ (spectral template fitting coefficient obtained
from principal component analysis of spectra) additionally ensures
that objects with spectra inconsistent with early-type galaxies are
not selected (note that eClass corresponds to negative $a$ in
\citealt{bern}). Finally, we ask that spectra have S/N~$>10$ in
$r$-band. Unlike \cite{bern} we do not require that the zWarning flag
not be set, nor do we impose a redshift cut at this stage. These
criteria yield 54950 candidates, or some 40\% of all DR1 galaxies
with spectra. For these galaxies we retrieve dereddened Petrosian
magnitudes and their errors, velocity dispersion measurements, and
various other parameters.  

For these galaxies, we measure Lick ${\rm
Mg}_2$ and D4000 indices (using the \citealt{balo99} definition), using 
a special purpose analysis pipeline (see \citealt{trem04} for details). 
Also, by measuring the residual nebular emission
lines left from the subtraction of a stellar continuum model \citep{jarle}
we can classify these galaxies using a BPT plot \citep{bpt}
into star-forming galaxies (3\%), star-forming galaxies with weak S/N
features (13\%), galaxies that show both star forming and AGN features
(composites, 8\%), and AGNs (19\%). The remainder comprises galaxies
with $S/N<2$ in H$\alpha$, i.e. no measurable lines permitting classification
in any of the above categories.  This is our core sample of
 ``quiescent'' early-type galaxies. They
represent 57\% of our sample (yet just 30\% of the full DR1 are
quiescent early-type galaxies).

GALEX images the sky in the far and near UV (1530 \AA\ and 2310 \AA)
in two modes: All-sky Imaging Survey (AIS, $m_{\rm lim}(\rm AB)\approx
20.4$) and Medium-deep Imaging Survey (MIS, $m_{\rm lim}(\rm
AB)\approx 22.7$). Each circular field covers 1.1 sq.\
deg. \cite{seibert} have performed SDSS DR1/GALEX matching of objects
in GALEX internal release (IR 0.2) catalog, consisting of 649 AIS and
94 MIS fields, of which 117 and 91 overlap (not always entirely) with
SDSS DR1.   We include an additional field (not in IR0.2) on the cluster of galaxies
Abell 2670 $(z=0.076)$.
Thus our total overlap with SDSS DR1 is
$\sim 144$ sq.\ deg., or some 10\% of DR1 spectroscopic footprint.  We
use $FUV$ and $NUV$ magnitudes and errors derived in an elliptical
aperture.  We cross-correlate our 54950 early-type candidates from DR1
with GALEX objects having SDSS spectroscopic counterparts (10357
objects) and obtain 2174 objects in common. 

We discard 120 galaxies with $z>0.2$ because of our compromised
ability to classify them, and due to the 2 Gyr lookback time.
We also exclude 24 objects not classified
spectroscopically as galaxies. This leaves us with 2038 galaxies
with acceptable SDSS and GALEX data, with morphology and basic spectral
shape consistent with early-type galaxies, and with $z<0.2$. We then
subject this sample to further scrutiny using the BPT classification
outlined above. Now the star forming (SF) galaxies constitute 6\%,
low-S/N SF 17\%, composites 15\%, AGNs 27\%, and the remaining,
quiescent early-types (QE), 35\% of the sample. Since GALEX is
sensitive to UV light from young stars and nuclear activity, it is not
surprising that the quiescent galaxies have fallen in proportion with
respect to SDSS-only sample. Although we initially
select all of these five classes 
as early-type candidates, many of the non-QE galaxies
(especially SF) are obviously late-types.  Their inclusion in the
candidate list is often times due to the presence of a
dominant bulge, with a relatively faint disk.

In our analysis we will primarily be concerned with galaxies that
exhibit FUV flux. Of our 2038 galaxies, only one-half (1032) have
detectable (${\rm S/N}>3$) FUV flux. It is these galaxies that are
plotted in Figure 1. However, in this FUV-selected sample, QE comprise
only 18\%, or 189 objects. These optically ``old red and dead'' galaxies are the focus
of our study.  Since, as we have seen, relying on just the
morphological criteria is unreliable, and further considering that
the fiber spectra (in contrast to the
integral fluxes) sample only the central $3^{\prime\prime}$ of a galaxy, we have
visually inspected SDSS images of these 189 galaxies.  As a result of
this, we find 17 objects that show hints of a structure: bars, weak
spiral arms, or rings.  While their classification as true early type galaxies
is in doubt, the colors of these objects are no bluer, nor their spectra suspect,
compared to our main QE sample.
We remove these objects from our final sample of
172 quiescent E/S0s with a detectable far-UV flux. Roughly 1/2 of
these galaxies lie closer than $z=0.1$ and the other half at
$0.1<z<0.2$.

We visually inspected
the spectra of the 172 QE galaxies, and find that they are homogeneous
with no abnormalities consistent
with the presence of a late burst of star formation not accompanied
by $\rm H\alpha$ emission.  The S/N of the QE sample spectra
is $23 \pm 7$ (rms); we are confident that we did not miss any
significant emission.  The lowest S/N quartile of our sample
has FUV$-r=4.5\pm 0.8$, while the highest S/N quartile
has FUV$-r=6.8\pm 0.7$.  We attribute this difference to bright red
nearby galaxies which drop out of the sample at $z>0.1$.

\section{Discussion}

We first consider the 
UV-optical properties of the full sample
of SDSS DR1 galaxies with early-type morphology, spectra with
S/N$>10$, and
detections in the FUV band (Figure 1).   
These plots
show a weak correlation between $\rm Mg_2$ and $FUV-r$, but
in the sense that more metal rich galaxies are redder, opposite the trend 
of BBBFL.    
Notably, galaxies with even weak spectroscopic signatures of star formation
tend to have the bluest FUV colors and uniformly lower $\rm Mg_2$,
and even lower D4000 (not shown).  
The
distribution of the AGN and especially of 
star forming galaxies look similar in this
plane, and clearly differs from that of the QE galaxies in the lowest panel, which
mostly have $\rm Mg_2 > 0.2$, and very few galaxies with $FUV-r <3.5$.
Averging $\rm Mg_2$ for galaxies with $FUV-r<6$, we find 0.16 for the
star forming, 0.22 for the AGN, and 0.24 for the QE.

Figure 2 concerns only possible correlations within the QE sample,
now separated by redshift.
We also identify
members of Abell 2670 which has also been
investigated by \citet{lee04} to study the evolution of the UV rising flux.
A rich cluster of galaxies with a large population of ellipticals, one
might expect the galaxies in Abell 2670 to be more uniform in their age
of formation, and less affected by residual star formation, than 
are the field galaxies.  We see no tendency to recover the BBBFL trends
among the Abell 2670 sample.   The A2670 members appear to be uniformly
more red than the field, both for the $z<0.1$ and $0.1<z<0.2$ subsets.
In addition to $\rm Mg_2$, we consider the metallicity sensitive D4000 
index (the continuum drop below the Ca HK lines is due to a large increase
in metal lines, and is a metallicity sensitive index) and with velocity 
dispersion (mass).   We find a weak correlation between $FUV-r$ and
metallicity or mass, but in the opposite sense to that of BBBFL: more massive,
metal rich galaxies have redder $FUV-r$ color.

Concerned by this failure to reproduce the BBBFL trends, we checked
that the Mg$_2$ and D4000 correlate as expected, and that $\log \sigma$ and
$\rm Mg_2$ correlate.   We inspect visually each FUV detection and find a handful of cases where the UV
emision is extended, even for galaxies in the QE sample; removal of these from our sample
has no effect.  Early type galaxies with $NUV-r<5$ are classified as residual
star forming galaxies by \citet{yi04} in their study of the UV color-magnitude
relation.  They argue that such a strong UVX is inconsistent with that
found in local samples (e.g. BBBFL).   We assert that our spectroscopic
pre-selection and observed behavior in the D4000 and Mg$_2$ vs $FUV-r$ plots
qualifies these as true UVX candidates.  Even if such galaxies
were excised from the sample,
our conclusions remain unaffected.

The $z<0.1$ sample of QE galaxies appears to
be redder in $FUV-r$ color than the $0.1<z<0.2$ sample.
We have considered
whether this could be due to a K correction by convolving the SDSS and
Galex bandpasses with the appropriate quiescent evolving red metal
rich population, constructed from the \cite{bc03} models.  
The K correction (single burst population formed 13 Gyr ago)
alone will cause an evolution of 0.3 mag to $z=0.1$ and
0.5 mag to $z=0.2$
(a single burst stellar population formed at $z>3$ is already
$\sim 10$ Gyr old at $z=0.2$).  The difference in the color distribution
may be related to rapid evolution of the UVX (as is evidently also seen
in \cite{lee04}).  We will consider this and other issues in greater detail,
later.  However, we emphasize that we do not recover the BBBFL trends
even for the lowest redshift subset of our data.  

Surprising in light of their spectroscopic uniformity is the  5 mag
range of $3.0 < FUV-r <8.2$.  While half of our initial sample of early
type galaxies was not detected in FUV, we verified that the bright
end of our sample is sufficiently well detected in the FUV that
the red limit is physical and not due to the reddest galaxies being
undetected.
The $FUV-NUV$ color may signal changes in the character of the underlying stellar population
responsible for the UV upturn.  Figure 3 suggests that galaxies
with the reddest $FUV-r$ (weakest UVX) also have redder $FUV-NUV$ colors.  
Note that the QE galaxies are bluer
than SF in $FUV-NUV$: only 3\% of the SF lack an NUV  detection, while
16\% of the QE galaxies lack NUV.

In light of our absence of a correlation between UVX and metallicity,
it becomes difficult to invoke helium abundance (entering
through $\Delta Y/ \Delta Z$) as the primary controlling parameter in the
strength of the UVX.
The wide variation in color and lack
of correlation with an obvious physical property might reflect
small variations in star formation history which affect the age 
distribution of the old stars, or with UVX arising from a rare population fed by
stochastic mass loss on the first ascent of the red giant branch.

The lookback time in our sample is not negligble.  For the concordance
cosmology of $\rm H_0=70\ km sec^{-1}Mpc^{-1}$, $\rm \Omega_M=0.3$, and $\Omega_\Lambda=0.7$,
the lookback time at $z=0.1$ is 1.3 Gyr and reaches 2.4 Gyr at $z=0.2$.
Small differences in the formation time of our QE sample could correspond to observed
diffferences of 1-2 Gyr in age.   Well populated blue horizontal branches are found only
in globular clusters with the age of the Galactic halo.   This observation is reflected
in theoretical models (e.g. \citealt{ yi99, lee02}) 
which predict that the FUV
flux should fade by 2-4 mag within the 2.4 Gyr lookback time spanned by this
sample. 

\galex continues its surveys, and we plan to exploit the overlap between these
and the SDSS DR2, resulting in a large jump in sample size.  Normal science operations with \galex did not begin until August 2003,
however, observations of the Virgo cluster have now been obtained and we
expect to be able to reproduce more closely the measurements of BBBFL both
for nuclear and integrated light, for nearby early-type galaxies.  The origin
of the UV rising flux and its systematic behavior with physical properties remains  a 
complicated and challenging problem, some 30 years after its discovery using
the first orbiting ultraviolet telescope.

\acknowledgments GALEX (Galaxy Evolution Explorer) is a NASA Small Explorer, launched in April 2003. We gratefully acknowledge NASA's support for construction, operation, and science analysis for the GALEX mission, developed in cooperation with the Centre National d'Etudes Spatiales of France and the Korean Ministry of Science and Technology.  We also acknowledge use of data from the Sloan Digital Sky Survey.
The authors thank T. Brown and A. Renzini for valuable comments.
JB acknowledges the support of an ESA postdoctoral fellowship.

\begin{figure}
\epsscale{0.7}
\plotone{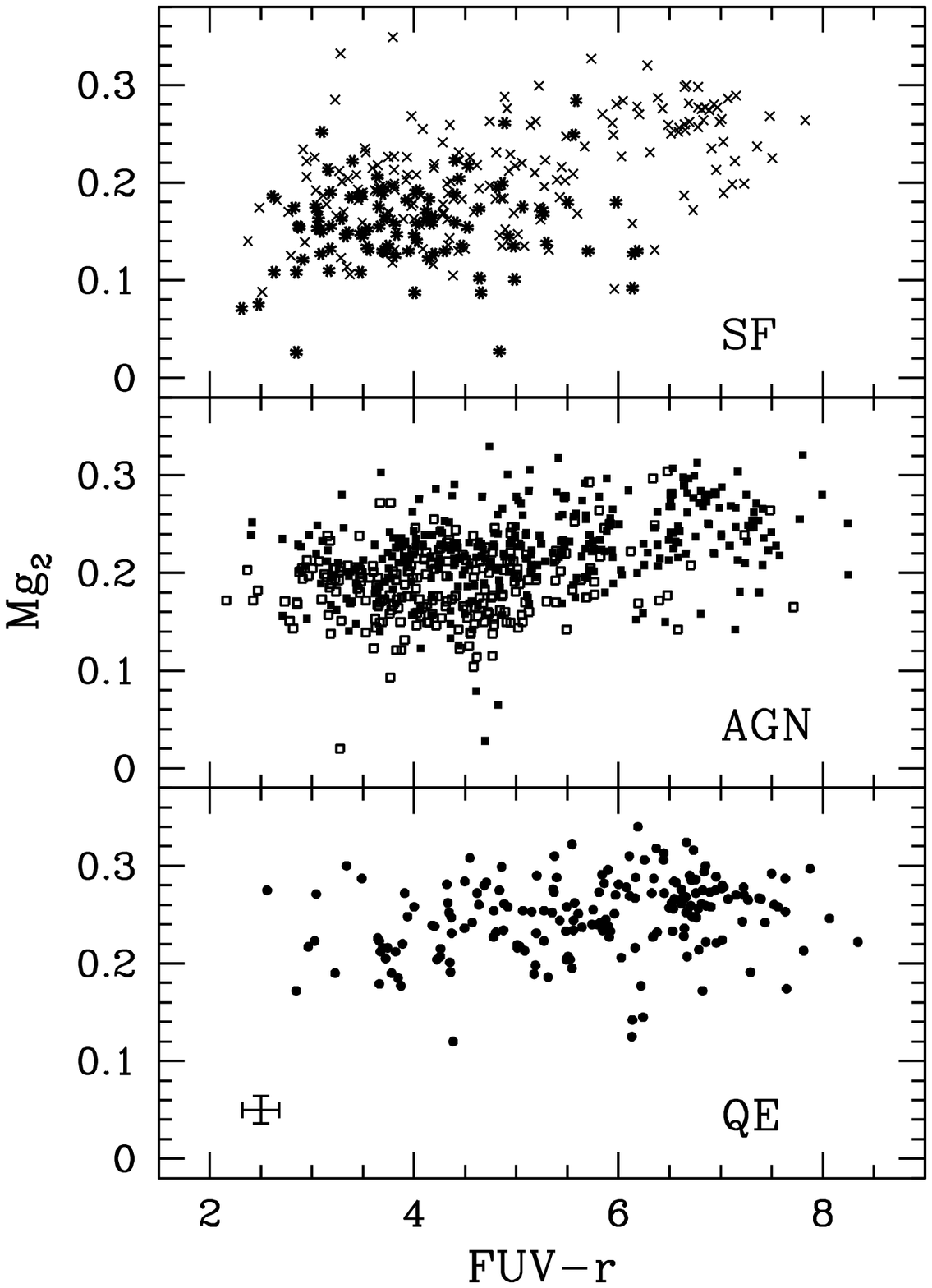}
\caption{Early type galaxies with both GALEX FUV and SDSS spectra which
have been classified according to their emission lines as star forming,
weak star forming, AGN, AGN/star forming composite, or quiescent;
$\rm Mg_2$ vs $FUV-r$.  Upper panel:  star forming galaxies.  Asterisks
indicate well detected star forming galaxies; crosses indicate
galaxies with star formation detected at low S/N.  A large fraction of these
galaxies have $FUV-r<5$ and $Mg_2 <0.2$.     Middle Panel:  Galaxies
with AGN.  Filled squares show pure AGN, while open squares are
star forming/AGN composites.  Lower Panel:  quiescent elliptical sample (label QE) with
no emission lines.  Notice that these galaxies have
stronger $\rm Mg_2$ index and $FUV-r > 3$.  We consider this sample
to comprise the true UVX population.  There is only a weak correlation
between the $\rm Mg_2$ index and color; no hint of the BBBFL correlation is
present.  The full range in $FUV-r$ is 5 mag.  Cross indicates
typical $1\sigma $ errors.} 
\end{figure}

\begin{figure}
\epsscale{0.7}
\plotone{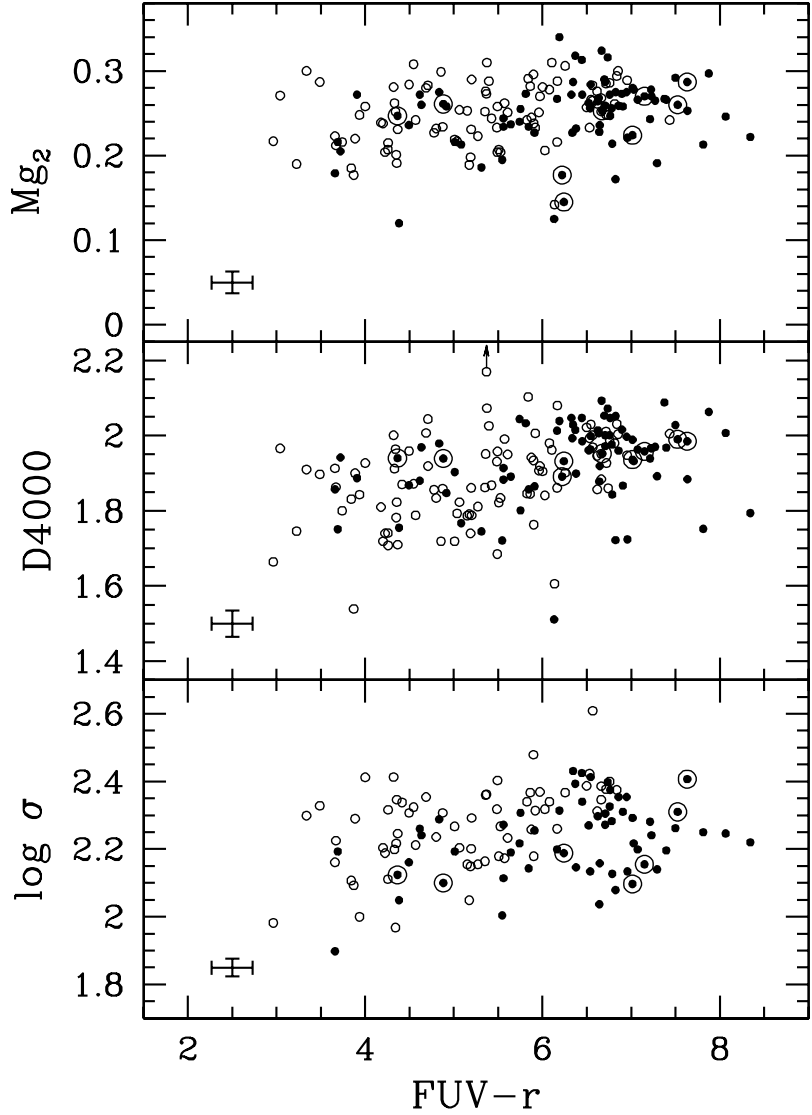}
\caption{Plot of various parameters vs $FUV-r$ for the 172 quiescent
elliptical galaxies in our sample.  Filled circles indicate $0<z<0.1$;
open circles indicate $0.1<z<0.2$.  Circled points indicate membership
in the elliptical rich cluster Abell 2670 $(z=0.076)$.    
These panels show that $\rm Mg_2$,
D4000, and $\log \sigma$ are all weakly correlated with $FUV-r$
color, in contradiction to the trends reported in BBBFL.   The
Abell 2670 members have weak UV rising flux and do not show any
tendency for the UVX to be higher in more metal rich galaxies.
One galaxy in the middle panel has D4000 = 2.5 and is indicated with
an arrow. The crosses indicates $1\sigma$ errors.}
\end{figure}

\begin{figure}
\epsscale{0.6}
\plotone{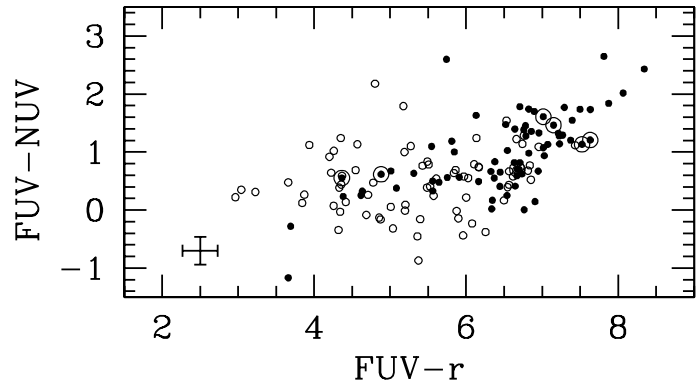}
\caption{Plot of $FUV-NUV$ vs $FUV-r$ for our sample of 172 quiescent 
early-type
galaxies.  Filled circles indicate $0<z<0.1$; open circles, $0.1<z<0.2$
Both colors tend to become redder for the reddest galaxies in the sample.
The galaxies with the strongest UVX do not have bluer $FUV-NUV$ colors.  This
suggests that the UVX, when present, does not have systematic dependence of
the slope on color. Cross indicates $1\sigma$ errors.}
\end {figure}


\begin{thebibliography}{}



\bibitem[Abazajian et al.(2003)]{dr1} Abazajian, K., et 
al.\ 2003, \aj, 126, 2081 

\bibitem[Baldwin, Phillips, \& Terlevich(1981)]{bpt}
Baldwin, J.A., Phillips, M.M., \& Terlevich, R. 1981, \pasp, 93 5

\bibitem[Bernardi et al.(2003)]{bern} Bernardi, M., et al.\ 
2003, \aj, 125, 1817 

\bibitem[Balogh et al.(1999)]{balo99}
Balogh, M.~L., Morris, S.~L., Yee, H.~K.~C., Carlberg, R.~G., \&
Ellingson, E. \ 1999, \apj, 527, 54


\bibitem[Brinchmann et al.(2004)]{jarle} Brinchmann, J., et al.\ 2004, 
\mnras, 351,1151


\bibitem[Brown et al.(1997)]{brown97} Brown, T.~M., Ferguson, H.~C., Davidsen, A.~F., \& Dorman, B.\ 1997, \apj, 482, 685 

\bibitem[Brown et al.(2000)]{brown00} Brown, T.~M., Bowers, C.~W., Kimble, R.~A., Sweigart, A.~V., \& Ferguson, H.~C.\ 2000, \apj, 532, 308 




\bibitem[Brown(2003)]{brown03}
Brown, T.M.  2003 , Ap\&SS, 291, 215

\bibitem[Bruzual \& Charlot(2003)]{bc03} Bruzual, G.~\& Charlot, S.\ 2003, \mnras, 344, 1000 

\bibitem[Burstein et al.(1988)]{b88}
Burstein, D. Bertola, F., Buson, L.M., Faber, S.M. \& Lauer, T.R. 1988, \apj, 328 440 (BBBFL)

\bibitem[Code(1969)]{code69}
Code, A.D. 1969, \pasp, 81, 475

\bibitem[Deharveng, Boselli \& Donas(2002)]{dehar02}
Deharveng, J.-M., Boselli, A., \& Donas, J.,  2002, \aa, 393, 843

\bibitem[Donas et al.(1987)]{donas87}
Donas, J., Deharveng, J.~M., Laget, M., Milliard, B., \& Huguenin, D.\ 1987, \aap, 180, 12

\bibitem[Gil de Paz(2004)]{gil04}
Gil de Paz, A. et al. 2004, \apj, in press [this volume]

\bibitem[Greggio \& Renzini(1990)]{greg90}
Greggio, L, \& Renzini, A. 1990, \apj, 364, 35

\bibitem[Greggio \& Renzini(1999)]{greg-ren99}
Greggio, L, \& Renzini, A. 1999, Mem.Soc.Astron.Ital., 70, 691

\bibitem[Lee, Lee, \& Gibson(2002)]{lee02} Lee, H., Lee, Y., \& Gibson, B.~K.\ 2002, \aj, 124, 2664 

\bibitem[Lee et al.(2004)]{lee04}
Lee,Y.-W. et al. 2004, \apj, in press [this volume]

\bibitem[Martin et al.(2004)]{martin04} 
Martin, C. L., et al. 2004, \apj, in press [this volume] 

\bibitem[Morrisey et al.(2004)]{mor04} 
Morrissey, P.  et al. 2004, \apj, in press [this volume] 

\bibitem[O'Connell(1999)]{oconnell99}
O'Connell, R.W. 1999, \araa, 37, 603

\bibitem[O'Connell et al.(1992)]{ocon92}
O'Connell, R.W. et al. 1992, \apj, 395, L45

\bibitem[Ohl et al.(1998)]{ohl98} Ohl, R.~G., et al.\ 1998, 
\apjl, 505, L11 


\bibitem[Rich et al.(1997)]{rich97}
Rich, R.M. et al. 1997, \apj, 484, L25


\bibitem[Seibert et al.(2004)]{seibert} Seibert, M., et al.\ 2004, 
\apjl, this volume


\bibitem[Tremonti et al.(2004)]{trem04}
Tremonti, C.A. et al. 2004, \apj, in press; astro-ph/0405537

\bibitem[Yi et al.(1999)]{yi99} Yi, S., Lee, Y., Woo, J., Park, J., Demarque, P., \& Oemler, A.~J.\ 1999, \apj, 513, 128 

\bibitem[Yi et al.(2004)]{yi04} 
Yi, S.K., et al. 2004, \apj, in press [this volume] 



\end{thebibliography}
\end{document}